\documentstyle[12pt]{article}

\begin{document}
\begin{titlepage}
\begin{center}

October 6, 2000     \hfill    LBNL-46871 \\

\vskip .5in

{\large \bf The importance of quantum decoherence in brain processes}
\footnote{This work is supported in part by the Director, Office of Science, 
Office of High Energy and Nuclear Physics, Division of High Energy Physics, 
of the U.S. Department of Energy under Contract DE-AC03-76SF00098}

\vskip .50in
Henry P. Stapp\\
{\em Lawrence Berkeley National Laboratory\\
      University of California\\
    Berkeley, California 94720}
\end{center}

\vskip .5in

\begin{abstract}
It is shown how environmental decoherence plays an essential and 
constructive role in a quantum mechanical theory of brain process 
that has significant explanatory power.
\end{abstract}

\end{titlepage}

\newpage
\renewcommand{\thepage}{\arabic{page}}
\setcounter{page}{1}

\noindent {\bf 1. Introduction}

In a recent article with the same title as this one 
Max Tegmark[1]  estimated how long it would take for 
interactions with the environment to destroy 
macroscopic quantum coherence in the brain. He 
arrived at the result $10^{-23}$ seconds. This tiny 
value appears, on the face of it, to rule out any 
significant role for macroscopic quantum effects 
in understanding the connection between brain 
processes and conscious thoughts. Indeed, Tegmark 
used his results to discredit Penrose's theory of 
consciousness. 

Tegtmark noted that I also had developed a detailed  
theory of the mind-brain connection, but he did not 
direct his remarks about decohence at my theory. 
That would have made no sense, for his results are 
constructive rather than destructive in the context 
my theory, which is based heavily on the presumption that 
enviromental decoherence has a large effect on brain 
processes. My theory is specifically designed so that 
the particular quantum effects that allow a person's thoughts 
to influence his brain are not affected by environmental 
decoherence. This stringent requirement imposes  
non-trivial conditions on human behavior under controlled
situations, and the empirical data gathered by psychologists
during the past fifty years indicate that these detailed
conditions are satisfied.

The foregoing remarks make it clear that the 
macroscopic quantum effects exploited in this quantum theory 
of the mind-brain system are {\it not} the macroscopic 
quantum effects used in quantum computation. Those latter 
effects are obliterated by environmental decoherence. Thus if 
brains really do operate in the way described by this theory 
then the technical ramifications could be far reaching: it 
would make available for development a type of macroscopic quantum 
effect that has been exploited by biological systems, but has
not been been used in engineering. No new principle is involved
here: the theory is simply a rational consequence of taking 
seriously the principles laid down by John von Neumann[2] 
and Eugene Wigner[3], together with reasonable and standard 
ideas about what brains do.

This theory has been described in a number of 
documents[4] designed to inform neuroscientists and 
psychologists, but has not been described in a physics journal, 
in a form directed at physicists. The purpose of this paper
to fill that void, and to show in particular how
macroscopic quantum effects can be effective 
in a system that is decomposed into a mixture of nearly 
classical states on a time scale of $10^{-23}$ seconds.
To establish that this theory is science, not mere speculative 
philosophy, I shall describe at the end some of the explanatory 
power of the theory. 

My central task here is to explain how macroscopic quantum 
effects occurring in the presence of massive environmental
decoherence can produce an action of mind on brain
that can account for the detailed empirical facts.
I assume that there is nothing special 
about a human brain except for its physical structure, 
and hence that other systems with similar 
functional structure should exhibit similar behaviour.

Many physicists are unfamiliar with the profound 
difference between the von Neumann/Wigner
formulation of quantum theory and the more common
Copenhagen formulation. So I will begin by
describing the elements of the von Neumann/Wigner theory,
before turning first to the central theoretical task, and 
then to confrontation with data.

\noindent {\bf  2. The General Theory.}

Bohr, Heisenberg, Dirac, Born, Pauli, and the other 
founders of quantum theory formulated their approach in a 
way that constituted a radical break with the classical 
physics that had preceeded it. They recognized, more 
explicitly than their immediate predecessors, that 
science was basically a human endeavour, and they
defined the proper objective of science to be the 
construction of rules that would allow human beings to 
make useful prediction about connections between their 
observations. Thus human experiences were elevated to the 
status of the primary reality dealt with by the theory. 
Physicists were enjoined to desist from all efforts to 
understand the reality that lies behind their observations. 
Classical concepts were brought in through the fact 
that our description of how we set up our experiments, 
and what we learn from them, are---as a matter of 
fact---couched in the ordinary language of everyday life, 
refined by the concepts of classical physics. 

This way of understanding quantum theory is called the 
Copenhagen interpretation, and  it is  very useful: 
it allows physicists to get on with the job of testing 
the rules and applying them, without getting embroiled
in the puzzling features that arise when one tries to
dig deeper.

Von Neumann, however, did dig deeper. He noted 
that the measuring devices, which had a rather strange 
dual status in the Copenhagen approach, since they were 
physical objects, yet were descibed in classical language,
were made up of the same kind of atomic constituents as 
the atomic systems that they were probing, and that 
there was therefore a well defined way to include these
devices into the system that was described by the 
mathematical formulas of quantum theory, provided one 
brought into this description the entire physical universe, 
including our bodies and brains. This mathematical description
was thereby elevated from its former status of being merely the
language for the formulation of a  mysterious set of rules for 
making predictions about our conscious experiences, to the status 
of a new kind of description of the physical universe. However,
this description, unlike its classical predecessor, was 
dynamically linked to our conscious experiences. This 
dynamical linkage between the aspects of nature that 
were described in this mathematical language and the aspects 
of nature that we describe as the contents of our streams of 
conscious thoughts allowed physicists to obtain from a single 
unified theory all the predictions of Copenhagen quantum theory 
and all the valid prediction of classical physical theory from 
a dynamical theory that describes also the interaction between 
our minds and our brains. The key dynamical interaction 
is a change, in conjunction with each conscious experience,
in the mathematically described universe: this change, called a 
`reduction' or a `collapse', tends to eliminate all those 
potentialities of the previously existing state that are incompatible 
with the expectations that characterize that experience. 

This ``reduction of the wave packet''  is the basis of 
Copenhagen quantum theory. But von Neumann/Wigner quantum theory
shifts the effect of the reduction associated with a person's 
experience away from some external system, such as the measured atomic 
system, or the external measuring device, onto that person's brain. 
This placement is, from a scientific point of view, a far more 
reasonable positioning of the action of mind on matter, at least  
if one is trying to construct a theory of what is really 
happening. This placement creates the basis for 
a dynamical theory of how our conscious thoughts influence 
our brains. On the other hand, no commitment need be made 
as to the underlying connections between the realities 
that we describe as elements of our streams of conscious
thoughts and the aspects of nature represented by the 
mathematical description of the physical world. For this 
theory is, in the final analysis, just a theory of the 
interplay between the aspects of nature that we describe 
in these two different ways, regardless of the true nature 
of the underlying realities. 

To express these ideas in mathematical form let $S(t)$ 
represent the state of the universe at time t. 
Actually, the variable t labels a sequence
Tomonaga-Schwinger[5] spacelike surfaces  $\sigma (t)$ 
such that the whole surface moves gradually forward  
as $t$ increases. I assume that for the study of brain 
dynamics quantum electro-dynamics will be adequate for 
the pertinent ranges of size and energies, and that 
future developments in elementary-particle physics will 
provide the necessary ultra-violet cut-off. I am 
imagining, for definiteness, that the surfaces 
$\sigma (t)$ will be the constant-time surfaces in the 
rest frame of the cosmic background radiation.

The state $S(t)$ is the {\it operator} form of the state:
$S(t)/Tr S(t)$ is the usual `density matrix', where Tr 
stands for the trace operation.

The operator $S$ can be represented by a matrix $S_{ij}$. 
If the system represented by $S_{ij}$ is composed of n 
component parts, then the index i will consist of a sequence 
of n indices, and j will also be represented in this way. 
If $b$ is the set of indices labelling some critical part  
of some person's brain, and $-b$ represents the complementary 
set of indices---i.e., the set of indices for rest of the 
universe---, then the state of this brain part, is represented
by $S(t)_b = Tr_{-b}S(t)$, which is the partial trace 
of $S(t)$ over the complementary set of variables $-b$. 
This operator $S(t)_b$  acts in the subspace of indices
$b$ associated with this brain part.

The basic problem to be faced is that the
interaction of the brain with its environment will keep
$S(t)_b$ in the form of a mixture of almost classical 
states, and that this would seem, on the face of it, to 
preclude macroscopic quantum effects of the kind needed 
for a quantum-mediated influence of mind on brain.

To see why that is not true one must understand what
the brain is doing.

The job of the brain is basically to take clues, coming 
via sensors, about the situation of the body in its 
environment, and construct an appropriate plan of action, 
and then to initiate and supervise the execution of this plan. 
I assume that the evolutionary process has honed the 
properties of the mind-brain so that it performs this task well.

How do quantum effects enter into the behaviour of
$S(t)_b$?

There is, of course, the basic fact that quantum theory 
is needed to make the chemistry work right. But in order
to get directly to the essential point let me grant that 
the chemical interactions could be mocked up by some
essentially classical-type model, and that the whole 
brain can be treated classically except for one thing: 
the migration of calcium ions  within nerve terminals 
from the exits of micro-channels to the sites where they
trigger the release into the synaptic cleft the contents 
of vesicles of neurotransmitter. 

The diameters of the micro-channels in cerebral nerve 
terminals are approximately one nm [6], This means that
the indeterminacy in the velocity of the migrating calcium 
ion that arises from the Heisenberg uncertainty principle 
is smaller than its  thermal velocity by a factor of about 300.
The distance between microchannel exit and trigger 
site is about 50nm [7]. Thus the uncertainty in the
location of the calcium ion when it reaches the trigger
site is of the order of size of the calcium ion itself.
This means that the classical conception of the brain is
inadequate in principle: quantum effects will generate a 
superposition of the classical state in which the 
neurotransmitter in the vesicle is released and the classical 
state in which this packet of neurotransmitter is not released. 
This superposition will quickly be reduced to a mixture. 
A similar bifurcation occurs at each active nerve  terminal. 
Hence the state $S(t)_b$ will necessarily evolve into a 
{\it mixture} of a huge number of states. Actually, a 
{\it continuum} of possible states will contribute, 
because each vesicle could be released a little earlier 
or a little  later, and this will produce a continuum of 
contributing possibilities.

This first step is already important, because it shows
that the idea that classical physics could give a 
deterministic answer to  how the brain evolves
is in principle wrong: that possibility is {\it strictly 
 incompatible} with quantum theory, even if one ignores 
quantum effects associated with chemistry. Quantum effects 
entail that the brain state $S(t)_b$ will quickly evolve 
onto a mixture of quasi-classical possibilities, all of 
which are actually present, insofar as no actual collapse 
has occurred. It is important in what follows that the 
interaction with the environment, although it reduces 
superpositions to mixtures does {it not} reduce the 
mixture of quasi classical possibilities to a single one 
of these possibilies: all states of the mixture will 
continue to exist in parallel, insofar as the 
evolution is controlled by the Schroedinger equation.

It is, of course, well known that if one computes
the expectation value of any operator in a state $S$
that is a mixture of states $S'$ then the result is  
identical to what would be obtained if the state were
really in just {\it one} of the states $S'$, but that
one does not know which state $S'$ this is, but does 
know the probability of each of the states $S'$. 

This fact might suggest that there could be no way to 
distinguish a quantum model of the brain from a classical 
statistical model. However, that conclusion is incorrect, 
within the von Neumann/Wigner framework.

Within this vN/W framework each alert person has a stream 
of conscious experiences, and each such experience is 
associated with a reduction of the state of his brain 
to a form that is compatible with the experience: 
the reduction eliminates from the state of the brain all 
patterns of activity that are incompatible with that 
experience.

This reduction is represented mathematically in the
following way. Each experience $E$ is associated with
a projection operator $P(E)$, and the occurrence of E
at time t has  the following effect: if the symbol
$t-$ signifies the action of taking the limit in which
the argument of $S(t')$ tends to time $t$ from times earlier 
than $t$, and $t+$ is related in the same way to 
later times, then
$$
S(t+)_b = P(E)S(t-)_bP(E).       \eqno (1)
$$
The  projection operator P(E) acts in the subspace
associated with the indices $b$, and satisfies
the defining condition for projection operators
$P(E)^2 = P(E)$. It acts on the brain state $S(t)_b$
to eliminate all patterns of activity or structure that 
are incompatible with E.

For example, if the experience E is an updating of 
the person's representation of the  world, then P(E) will 
preserve in the mixture of quasi-classical states $S'$ 
represented by $S(t)_b$ just those that contain  the
patterns of brain activity that will tend to etch into
memory the experienced updating. If the experience is of an 
intention to cause one's body to move in a certain way 
then P(E) will eliminate from the mixture $S(t)_b$ those
quasi-classical states $S'$ that do not have the patterns
of brain activity that will tend to cause the body to move 
in this way. These rules are just the analogs in the
vN/W framework of the Copenhagen rule that the
reduction is to the state that is compatible with the
experience. But in the vN/W framework it may be supposed
that this linkage of each experience $E$ with a $P(E)$
that tends to produce the expected or intended future 
experiences is a consequence of the evolutionary 
development of the human system.

In any attempt to go beyond the Copenhagen interpretation
of quantum theory the chief problem is the so-called
``Basis Problem'': What determines which {\it basis} will 
be used to reduce the myriads of possibilities produced by
the quantum uncertainties to the individual reality
that is experienced? Environmental decoherence is helpful,
but it is not sufficient, because the quasi-classical
states are over-complete, and hence do not provide
a unique basis of normalizable states, and the structure
of conscious experience is tied to the formation of
quasi-stable and accessible memories, as Zurek emphasizes
in his excellent reviews [8]. 

Copenhagen quantum theory resolves the basis problem 
in a simple way: ``The Observer'', who stands outside 
the Hilbert space structure, decides how he will set up 
the experiments, and this decision fixes the ``basis''. 
This means that in the Copenhagen interpretation it is 
the `free choice' on the part of the experimenter as to 
what he is interested in that fixes the basis. 

The vN/W solution is essentially the same as the 
Copenhagen one: the basis is fixed by the experience of 
``the observer''. Something beyond the Schroedinger evolution
is needed to fix the basis, and this is taken to be ``experience''.

But then what fixes experience? How is E determined? 
And how does `free choice' enter? The experience E 
cannot just pop out of nothing: it must be determined 
largely by the brain.  Any adequate dynamical theory of 
the mind-brain must explain not only how mind effects
brain, but also how brain affects mind.

The job of the mind and brain, acting together, is
to determine the best course of action in the
circumstances in which the mind-brain finds itself.
The brain is busy grinding out, via the Schroedinger 
equation, the host of possibilities
represented by the mixture $S(t)_b$. The `best' option
should be the one such that $P(E)$ has the greatest statistical
weight. Let E(t) be the E that  maximizes $Tr S(t)_b P(E)$.
This should be the best candidate for E at time t.

But the experiential events in the person's stream of 
consciousness occur only at discrete times, not continuously. 
Thus the `free choice' can be reduced to {\it consent}, 
at certain instants t, to put to Nature the question of 
whether experience E(t) will occur. If E(t) does occur 
then $S(t+)$ becomes $P'(E(t))S(t-)P'(E(t))$, in accordance
with (1). If Nature's answer is No, and hence E(t) does not 
occur at time $t$, then $S(t)$ becomes
$(1-P'(E(t)))S(t-)(1-P'(E(t)))$, where the operator $P'(E(t))$ 
in these expressions is the trivial extension to 
the entire space of the operator $P(E(t))$ defined previously.
The probability that the experience E(t) will occur is 
given by 
$$
Tr S(t)P'(E(t))/ Tr S(t) = Tr_b S(t)_b P(E(t))/ Tr_b S(t)_b. \eqno (2) 
$$
in accordance with the the basic probabilty rule of 
quantum theory.  

The point of all this is that Copenhagen quantum theory 
introduced our conscious experiences directly into physical 
theory, and von Neumann/Wigner quantum theory tied these 
experiences to projection operators that act on the person's 
brain state in such a way as  to bring the brain state
into concordance with the experience. This converts 
quantum theory basically into a dynamical theory of the
evolution of the universe that includes a theory of 
mind-brain interaction. 

In this theory behaviour is largely controlled by
the local mechanical brain process governed by the
Schroeodinger equation. That process is `local':
the interactions are basically contact interactions.
But there is one element that not governed by any known
law of physics, namely the choices to consent or not consent
at time t to putting to nature the question associated with
the possible experience $E(t)$. 

I do not intend to speculate at this point about how the 
evaluation that lies behind this choice is carried out. 
At the present early stage in the development of the science 
of the mind-brain system that question remains a project for 
future research. But the effect of a consent to `put the 
question to nature', is to force Nature to return an answer,
Yes or No, and the effect of the answer `Yes', is to activate 
the process (2), which tends to produce to updating or action
that characterizes the experience. This ``reduction'' of
``collapse'' process is macroscopic, in the sense that, according 
to the theory, the operator $P(E(t))$ acts instantaneously on 
some large part of the brain in accordance with process (1)
descrbed above. I presume that the evaluation process
has been honed by evolution so that it reflects the 
the likely consequences of activating process (1).

But in order for evolution to be able to hone this connection
it is necessary that one's choice to consent has a likely 
effect on behaviour.

In order to compute the {\it likely} effects on behaviour one must
add the properly weighted contributions from the two possible
answers that nature might give. This is exactly what
the famous von Neumann process 1 achieves: it gives the state
that represents the effect of putting to nature the question
$P=P(E(t))$ if no account is taken of which of  the two possible
answers, `Yes' or `No', nature returns.

The von Neumann process 1 is:
$$
S(t)_b=PS(t-)_bP + (1-P)S(t-)_b(1-P).     \eqno (3)
$$

So the key demand on the theory is that behaviour be controllable 
at the statistical level by applications at various times t of the 
process (3) associated with one's choice to give at time t the consent 
associated with the possible experience $E(t)$.  And this control
should tend to produce behaviour that conforms to the expectations 
and intentions imbedded in $E(t)$

Can one achieve this in a brain that is subject to massive
effects of environmental decoherence? That is the question.
The answer is Yes.

In order to consciously control one's behaviour one must normally 
keep attention focussed on a task for some period of time.
Think of the focus of attention  required to lift a heavy rock, 
or the focus of attention needed to fix into memory the details 
of some visual scene. In these clear examples of the
apparent control of brain process by conscious effort some 
particular thought, or aspect of thought, remains 
fixed and stable the mind-brain for an experienced  period of 
time. Hence we are led to consider the effect of putting the 
{\it same} question repeatedly to Nature in rapid succession.

Suppose, for definiteness, that the subsystem $b$ is a set of 
degrees of freedom that is generating a contribution to the 
low frequency $(< 40 Hz)$ part of the coulomb part of the elecromagnetic 
field in the brain, and that the von Neuman process (3) repeats 
rapidly on this scale.

Let $d$ represent the time interval between successive
actions of process (3). Then
$$
S(t+d)_b = 
$$
$$
P(\exp -iHd)[PS(t-)_bP+(1-P)S(t-)_b(1-P)](\exp +iHd)P
$$
$$
+
$$
$$
(1-P)
$$
$$
(\exp -iHd)[PS(t-)_bP+(1-P)S(t-)_b(1-P)](\exp +iHd)
$$
$$
(1-P).         \eqno (4)
$$

Because $Hd$ is small in the subspace associated with
$b$ one can approximate the exponentials by
$(1\mp iHd)$, and observe that the terms linear in
$d$ drop out: the effect of a rapid repetition
of process (3) is to damp out transitions between
the two subspaces specified by $P$ and $(1-P)$.
This will be recognized as the familiar Quantum 
Zeno Effect [9]. 

What this means is that if a person can, by willful effort, 
acting through his power to consent, increase the rapidity of the 
events in his stream of consciousness then he could control 
the activity of his brain by keeping the activity of the b 
part of it confined to the subspace it is already in. The 
brain state would be prevented from  ``wandering'' in the 
way that it would if there were no rapid quantum process (3).
Thus willful effort would alter the behaviour of 
the quantum brain, at the statistical level, from what it would
be if there were no macroscopic quantum effects.  

Note that the Quantum Zeno Effect described above is not
destroyed by the fact that $S(t)_b$ is a mixture: 
that makes no difference at all. The reason that the
macroscopic quantum effect persists in the presence of decoherence
is that it originates not in interference effects but rather
the fact that it is the whole brain part associated with the set of
variables b that enters into the dynamics, both at the level of
specifying the `best' possibility $P(E(t))$, and the associated
action (3).  

The key point is that the quantum theory of mind-brain 
described here gives each mind-brain the power, through force 
of will acting via consent, to keep its attention 
focussed on a task in a way that is impossible in the 
analogous classical statistical model. And this effect is 
achieved by increasing, through effort of will, the rate of 
events in the stream of consciousness.

I stress again that this theory, although it is more congenial
to ontological interpretation than the Copenhagen account,
is  basically just a way of organizing empirical features of the
study of mind-brain systems in a way that is suggested by, and
strictly compatible with, the basic laws and principles of physics.

\noindent {\bf 3. Explanatory Power}

Does this theory explain anything?

This theory was already in place [4] when a colleague brought to my attention
some passages from ``Psychology: The Briefer Course'', written by William 
James [25]. In the final section of the chapter on Attention James writes:

``I have spoken as if our attention were wholly 
determined by neural conditions. I believe that the array of {\it things}
we can attend to is so determined. No object can {\it catch} our attention
except by the neural machinery. But the {\it amount} of the attention which
an object receives after it has caught our attention is another question.
It often takes effort to keep mind upon it. We feel that we can make more 
or less of the effort as we choose. If this feeling be not deceptive, 
if our effort be a spiritual force, and an indeterminate one, then of 
course it contributes coequally with the cerebral conditions to the result.
Though it introduce no new idea, it will deepen and prolong the stay in 
consciousness of innumerable ideas which else would fade more quickly
away. The delay thus gained might not be more than a second in duration---
but that second may be critical; for in the rising and falling 
considerations in the mind, where two associated systems of them are
nearly in equilibrium it is often a matter of but a second more or 
less of attention at the outset, whether one system shall gain force to
occupy the field and develop itself and exclude the other, or be excluded 
itself by the other. When developed it may make us act, and that act may 
seal our doom. When we come to the chapter on the Will we shall see that 
the whole drama of the voluntary life hinges on the attention, slightly 
more or slightly less, which rival motor ideas may receive. ...''  
 
In the chapter on Will, in the
section entitled ``Volitional effort is effort of  attention'' 
James writes:

``Thus  we find that {\it we reach the  heart  of our inquiry  into volition
when we ask by what process is it that the thought of any given action
comes to prevail stably in the mind.}''

and later

``{\it  The essential achievement of the will, in short, when it is most 
`voluntary,'  is to attend to a difficult  object and hold it fast before
the  mind.   ...  Effort of attention is  thus the essential phenomenon
of will.''}

Still  later, James says:

{\it  ``Consent to the idea's undivided presence, this is effort's sole 
achievement.''} ...``Everywhere, then, the function  of effort is the same:
to keep affirming and adopting the thought  which,  if left to  itself, would 
slip away.''
  
This description of the effect of mind on the course of mind-brain process 
is remarkably in line with what arose independently from a purely 
theoretical consideration of the quantum physics of this process. 
The basic features of the interplay between effort, attention,
and control in the mind-brain system, as discerned by James, seem to 
come naturally out of the principles that von Neumann and Wigner
promulgated in their effort to make physical sense of the mathematical 
rules that explain the connections between the phenomena from the realm of
atomic physics. This opens up the interesting theoretical possibility of
bringing the whole range of science from atomic physics to mind-brain 
dynamics together in a single rationally coherent theory of an evolving 
physical reality made essentially of objective knowledge or information
rather than classically conceived matter.

A great deal has happened in psychology since the time of William James. 
A large amount empirical work pertaining to the issues at hand has been 
described in the book ``The Psychology of Attention'' by William Pashler [10].
This empirical work is basically behavioural: subjects are assigned
multiple tasks of various kinds and various loads, and their performances
are measured. Pashler makes a powerful case for the conclusion that
brain process is has two distinguishable subprocesses, one
more analytical and perceptual and operating via parallel processing, 
the other more selective of action, and acting via a linear ``bottleneck'' 
process. The detailed features of these two processes appear to be well
explained by the quantum model of the combined mind-brain system
developed in this paper, with the parallel processing aspect being
governed by the Schroedinger equation, and the selection bottleneck
being govern by the collapse process. 

I am not claiming that no classical model could explain these features.
But the fact that the details seem to be forced in the quantum approach
by the severe constraints imposed by the existence of strong environmental 
decoherence makes the quantum model more theoretically attractive than a 
classical model that puts these features, ad hoc, into a theoretical
structure in which consciousness can make no difference.

This journal is not the appropriate place to describe
the detailed ways in which the empirical findings described 
by Pashler support the  theory described above. A 
discussion of this matter can be found elsewhere [11].\\

\noindent {\bf References}\\

1. Max Tegmark, Phys. Rev. {\bf 61E}, 4194 (2000)

2. John von Neumann, {\it Mathematical Foundations of Quantum Mechanics}\\
   Princeton Univ. Press, Princeton NJ, 1955\\
   (Translation of 1932 Original)

3. Eugene Wigner, ``The Problem of Measurement'' and ``The Mind-Body Problem''
in {\it Symmetries and Reflections}, Indiana Univ. Press, Bloomfield, 1967.

4. Henry P. Stapp ``Attention, Intention, and Will in Quantum Physics''\\
   in {\it The Volitional Brain}, Benjamin Libet, Anthony Freemann, and\\
   Keith Sutherland, eds, Imprint Academic, 1999;\\
   ``Quantum Ontology and Mind-Matter Synthesis''\\
   in ``Quantum Future: Proceedings of the Xth Max Born Symposium''\\
   Springer-Verlag, Berlin, New York, 1999. \\
   See also http://www-physics.lbl.gov/\~{}stapp/stappfiles.html

5. S. Tomonaga, Prog. Theor. Phys. {\bf 1}, 27 (1946);\\
   J. Schwinger,  Phys. Rev. {\bf 82},  914 (1951).

6. E. Conley, {\it The Ion Channel Factsbook}, Acad. London, Vol IV, p.99.\\
   B. Hille, {\it Ionic Channels of Excitable Membranes},\\
   Sinauer Associates, Sunderland, MA, p.359.     
 
7. A. Fogelson and R. Zucker, Biophys. J., {\bf 48}, 1003 (1985).

8. W. Zurek, Prog. Theor. Phys. {\bf 89}, 281 (1993);\\
    and Phil. Trans. R.Soc. Lond. {\bf A 356} 1793 (1998).

9. W. Itano, D. Heinzen, J. Bollinger, and D. Wineland,\\
   Phys. Rev. {\bf 41A}, 2295 (1990).

10. Harold Pashler, {\it The Psychology of Attention}, \\ 
    MIT Press, Cambridge MA (1998)

11. Henry P, Stapp, ``From Quantum Nonlocality to Mind-Brain Process'',
    Lawrence Berkeley National Laboratory Report LBNL-44712.\\
    Submitted to Proc. Roy. Soc.\\    
    http://xxx.lanl.gov/abs/quant-ph/0009062\\
    and http://www-physics.lbl.gov/\~{}stapp/stappfiles.html
\end{document}